# Pulsed DSC Observation of the Static Spin Correlations

M.M. Nadareishvili[1], K.A. Kvavadze[1], N.E. Hussey[2], A.J. Shengelaya[3], S.J.Tsakadze[1], J.Ramsden[4]

1. E.Andronikashili Institute of Physics, Tbilisi State University, 6 Tamarashvili str., Tbilisi 0177, Georgia
2. University of Bristol, Tindal Avenue, bristol BS81TL, UK
3. Departament of Physics, Tbilisi State University, 2 Chavchavadze Av., Tbilisi 0179, Georgia
4. Cranfield University, UK

E-mail: m.nadareishvili@aiphysics.ge;  mnadarei@gmail.com

PACS: 74.72.–h Cuprate superconductors;
      74.25.Bt Thermodynamic properties.

**Key words:** differential calorimetry, LSCO, HTSC magnetic properties, static spin correlations.

**Abstract**

The investigations of low-temperature heat capacity in pure and Zn-doped $La_{1.84}Sr_{0.16}Cu_{1-y}Zn_yO_4$ samples ($y=0, y=0.033$) have been performed in the interval of 1.8–60 K by high-precision pulsed differential scanning calorimeter (PDSC), providing the measurements of heat capacities under the thermodynamically equilibrium conditions in contrast to commonly used differential scanning calorimeters (DSC). The anomaly of low-temperature heat capacity of a wide peak form related with Zn impurity induced static spin correlations was observed.

**Introduction**

The influence of magnetic and nonmagnetic impurities in the high-temperature superconducting cuprates $La_{2-x}Sr_xCuO_4$ (LSCO) have been extensively studied [1-3]. One of the most surprising features of impurity effects is doping LSCO with Zn that acts as a nonmagnetic impurity and is responsible for a strong decrease in the superconducting transition temperature Tc (~ -10K/%Zn). This effect is extraordinary large compared to that caused by substitution with magnetic impurities such as Ni [4]. On the other hand the neutron scattering experiments [5,6] show, that below 22K on the monocrystalline sample $La_{1.79}Sr_{0.21}Cu_{0.99}Zn_{0.01}O_4$ at the state with Zn suppressed superconductivity there appear elastic peaks suggesting that Zn influences on the AF dynamic spin correlations and stabilizes them into the static ones which have a spin-glass-like nature. At the same time, these experiments show that in the Zn-free LSCO static spin correlations are not observed [6]. The data of calorimetric investigations of this phenomenon are not available. In our opinion, the reason for this is insufficient precision of current DSCs.

The thermodynamic investigations of Zn-doped LSCO cuprates were carried out both with the usual pulsed calorimetric technique measuring the absolute heat capacity of samples [7,8] and with the DSC measuring the heat capacities difference between investigated and reference samples [9,10].

The high precision investigations of LSCOs heat capacity by conventional heat pulse technique under the equilibrium conditions [7,8] were carried out at the temperatures below 10 K, as above 10 K the phonon contribution to the heat capacity of each sample becomes large, consequently the absolute error of measurements rises.

The DSC investigations of HTSC cuprates [9,10] were used for the aim to eliminate large phonon contribution of investigated sample with corresponding contribution of reference sample at measurement heat capacities difference of these samples. But heat capacity anomaly connected with static spin corelations doesn't reveal itself. It should be noted that in DSCs a continuous heating of samples is used and, hence, the measurements are made under the conditions being non-equilibrium. Because at higher sensitivity of DSCs the precision of this measurement is not high.

## 2. Experiment

We have performed the measurement of low temperature heat capacity on the pure and Zn doped LSCO samples by the original method of pulsed differential scanning calorimeter (PDSC). At present, the most sensitive and, hence, the most commonly used calorimeters are the differential scanning calorimeters (DSCs) [11,12] including so-called Temperature Modulated DSCs (TM DSC) [13]. However these devices have one significant disadvantage - they measure the difference in heat capacity between the studied sample and the reference sample under the non-equilibrium conditions, as the heat capacity difference is measures in the continuous heating regime. For this reason they have comparativelly low precision and the measured physical characteristics often depend on the scanning rate [14].

Authors have developed a new method of calorimetry, on the basis of which pulsed differential calorimeter (PDC) possessing both high sensitivity and precision was created. The short description of this device was presented in [18]. This calorimeter combines the high sensitivity of the modern differential calorimeters of continuous heating type, and the high accuracy of the classical pulsed calorimeters. By this device the high precision measurements of the anomaly of low-temperature heat capacity were made in different substances [15,16], but unfortunatelly, this device was operated only at very low temperatures (T<30K), when the equilibrium establishment time for differential container, consisting of cells with sample and standard connected with thermo battery, is short. At the increase of temperature above 30K, the relaxation time of the differential container $\tau$ and, consequently, the time of pulsed experiments increases sharply due to the strong increase of heat capacity of differential container with samples ($\tau =C/k$ [17], where C is heat capacity of samples and k is heat conductivity of heat link between the cells through the thermo battery). This results in the decrease of the precision of pulse measurements and the creation of inconveniences at carrying out experiments. It became clear that the problem could be solved only partially in the first approximation. The way out was found in the development of the special new method of pulsed experiments allowing one to reduce the relaxation time of differential container to the relaxation time of one sample [18]. Implementation of this method of pulsed measurements gave authors capability to built the differential calorimeter of new generation – Pulsed Differential Scanning Calorimeter (PDSC) possessing simultaneously a high sensitivity and precision in a wide range of temperatures 1,5K – 400K.

In the pulsed regime, PDSC measures the difference in heat capacities $\Delta C$ between the studied and the reference samples under the equilibrium conditions, this difference makes a very small (~1%) part of the heat capacity of the sample and is within the error of measurements of the calorimeters measuring the absolute heat capacities. The relative error $\delta C/C \approx 10^{-4}$, where $\delta C$ is the error of measurement of $\Delta C$, and C is the absolute heat capacity of the sample.

The calorimeter also is capable to operate in continuous heating mode like the usual DSCs. The sensitivity of the device in the continuous heating regime is depend on the temperature and is~ $(7 \cdot 10^{-8} - 5 \cdot 10^{-9})$ W/10nv, the rate of scanning at the continuous heating is possible to change from 6 K/h to 60 K/h.

The samples $La_{1.84}Sr_{0.16}Cu_{1-y}Zn_yO_4$ where fabricated by a standard solid-state reaction and atter this they were sintered at the temperature of 1030 K while being subjected to a high pressure. Subsequent measurements of the susceptibility showed that the samples with x=0.16, y=0.033 did not display the superconducting transition for temperatures down to 4.2 K, being the lowest temperature in susceptibility measurements. On the other hand, the sample with x=0.16, y=0.00 showed the superconducting transition at the temperature of 38 K. The samples had the cylindrical form, the diameter ~20mm and the thickness ~3mm, the mass of each sample was ~4g.

## 3. Results and discussion

The heat capacity $C(y,T)$ of $La_{1.84}Sr_{0.16}Cu_{1-y}Zn_yO_4$ systems is the sum of electron (hole) $C_{el}(y,T)$ and phonon $C_{ph}(y,T)$ contributions:

$$C(y,T) = \gamma(y)T + C_S(y,T) + C_{ph}(y,T) \qquad (1)$$

where $\gamma(y)$ is the coefficient of linear term of $C_{el}$, $C_S(y,T)$ is the heat capacity of superconducting electrons.

Difference in heat capacity between the sample under investigation and the reference sample with the different Zn amount $\Delta C(y_1,y_2,T)$ is equal to

$$\Delta C(y_2,y_1,T) = \Delta\gamma(y_2,y_1)T + \Delta C_S(y_2,y_1,T) + \Delta C_{ph}(y_2,y_1,T) \qquad (2)$$

where
$\Delta\gamma(y_2,y_1) = \gamma(y_2) - \gamma(y_1)$
$\Delta C_S(y_2,y_1,T) = C_S(y_2,T) - C_S(y_1,T)$
$\Delta C_{ph}(y_2,y_1,T) = C_{ph}(y_2,T) - C_{ph}(y_1,T)$

Substitution of Zn for Cu does not affect the phonon heat capacity, as the masses and ionic radiuses of Zn and Cu are equal, and thus we can consider that $C_{ph}(y_2,T) - C_{ph}(y_1,T) = 0$ [7,8]. If the sample under investigation is superconducting with $y_2 = 0$ and the reference sample is non-superconducting with $y_1 = y_c$ ($y_c = 0.033$ is the minimal concentration of Zn, when superconductivity is suppressed), i.e. $C_S(y_1,T) = 0$, Expression 2 gives

$$\Delta C(0,y_1,T) - \Delta\gamma(0,y_1)T = C_S(0,T) \qquad (3)$$

Using the high-precision PDSC technique, the heat capacity difference $\Delta C(0,y_1,T)$ between the superconducting sample under investigation $La_{1.84}Sr_{0.16}CuO_4$ ($y=0$) and the non-superconducting reference sample $La_{1.84}Sr_{0.16}Cu_{0.967}Zn_{0.033}O_4$ ($y=0.033$) was measured in the low temperature interval of 1.8 – 50K under the equilibrium conditions. Fig.1 shows the $\Delta C(0,y_1,T) - \Delta\gamma(0,y_1)T$ ($=C_S(0,T)$) plot of this investigation. The difference between the coefficients of linear terms $\Delta\gamma(0,y_1)$ was estimated in a usual way, using the Debue model, by plotting the generally used relation $\Delta C(0,y_1,T)/T = f(T^2)$ [7,8]. On the basis of the $\Delta C(0,y_1,T)$ experimental data it was found that $|\Delta\gamma(0,y_1)| = 8.4$ mj/mol.$K^2$.

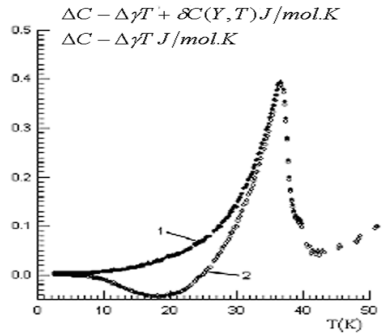

Fig.1
$\Delta C(0,y_1,T) - \Delta\gamma(0,y_1)T = f(T)$ dependence for molar heat capacities difference between $La_{1.84}Sr_{0.16}Cu_{0.967}Zn_{0.033}O_4$ and $La_{1.84}Sr_{0.16}CuO_4$

Curve 2 in Fig.1 (open circles) shows $\Delta C(0,y_1,T) - \Delta\gamma(0,y_1)T$ ($=C_S(0,T)$) dependence for molar heat capacity difference between $La_{1.84}Sr_{0.16}CuO_4$ and $La_{1.84}Sr_{0.16}Cu_{0.967}Zn_{0.033}O_4$

($y_1=0.033$ reference sample). One can note the appearance of the unphysical (negative) region for „$C_S(0,T)$".

This means that in the non-superconducting Zn-doped LSCO cuprate the anomaly of low temperature heat capacity appears, which has the form of a wide peak in agreement with neutron scattering experiments.

Thus, it is evident that the representation of the heat capacity of Zn-doped LSCO in the form of sum (1) is not complete and needs some correction, namely, the introduction of excess $\delta C(y,T)$ contribution.

$$C(y,T) = \gamma(y)T + C_S(y,T) + C_{ph}(y,T) + \delta C(y,T) \qquad (4)$$

The anomaly of heat capacity $\delta C(y,T)$ is related to the heat capacity of spin-glass-like state (static spin correlations) induced by Zn impurity observed in neutron scattering experiments.

Taking into account the additional term in heat capacity (exp.4), we obtain for $\Delta C - \Delta \gamma T$ the following expression

$$\Delta C(0,y_1,T) - \Delta\gamma(0,y_1)T = C_S(0,T) - \delta C(y_1,T) \qquad (5)$$

which explains the reason of appearance of the negative region on Fig.1, curve 2.

Fig.1, curve 1 shows the dependence $C_S(0,T) = \Delta C(0,y_1,T) - \Delta\gamma(0,y_1)T + \delta C(y_1,T)$ with taking into account contribution from static spin correlations),

**Conclusion**

Summing up the above said, one can conclude that in LSCO ceramic superconductors, at the introduction of Zn in nonsupeconducting state the anomaly of low-temperature heat capacity related with Zn induced static spin correlations was observed, according to neutron scattering experiments.